\documentclass[structabstract]{aa}  

\usepackage[dvips]{graphicx}
\usepackage{txfonts}
\usepackage{natbib}
\usepackage{hyperref}
\usepackage{multirow}
\begin{document}

\title{Effect of metallicity on the gravitational-wave signal from the cosmological population of compact binary coalescences}

\author{I. Kowalska-Leszczynska
         \inst{1}
         \and
         T. Regimbau\inst{2}
         \and
         T. Bulik\inst{1}
         \and
         M. Dominik\inst{1}
         \and
         K. Belczynski\inst{1}
         }

  \institute{Astronomical Observatory, University of Warsaw, Al Ujazdowskie 4,
00-478 Warsaw, Poland
\and UMR ARTEMIS, CNRS, University of Nice Sophia-Antipolis, Observatoire de la C\^{o}te d'Azur, CS 34229 F-06304 NICE, France
}      

  \date{Received  ; accepted  }

%%%%%%%%%%%%%%%%%%%%%%%%%%%%%%%%%%%%%%
\abstract
  {Recent studies on stellar evolution have shown that the properties of
compact objects strongly depend on the metallicity of the environment in which they were formed.
 }
  { Using some very simple assumptions on the metallicity of the stellar populations,
we explore how this property affects the unresolved gravitational-wave
background from extragalactic compact binaries.}
  {We obtained a suit of models using population
synthesis code, estimated the gravitational-wave background they produce, and discuss its detectability with second- (advanced LIGO, advanced Virgo) and third- (Einstein Telescope)
generation detectors.}
  {Our results show that the background is dominated by binary black holes for all considered models in the frequency range of terrestrial detectors, and that it could be detected in most cases by advanced LIGO/Virgo, and with Einstein Telescope with a very high signal-to-noise ratio. The observed peak in a gravitational wave spectrum depends on the metallicity of the stellar population.}
  {}

%%%%%%%%%

%We find the same structure, a spectrum that evolves as $f^{2/3}$ up to $40-100$Hz, corresponding to the inspiral phase, then as $f^{5/3}$ up to $100-500$ Hz, corresponding to the merger phase, before it reaches a maximum and decreases dramatically.  
%%%%%%%%%


\titlerunning{Effect of metallicity on the gravitational-wave signal}

\maketitle

\section{Introduction}
The coalescence of two neutron stars (NSNS), two black holes (BHBH), or a neutron star and a black hole (BHNS) are the most promising sources of gravitational waves (GWs) for terrestrial detectors, because of the huge amount of energy emitted in the last phase of their inspiralling trajectory, the merger, and the ringdown.
The coalescence occurs after two massive stars in a binary system have burned all their nuclear fuel, have evolved into red giants, and the cores have collapsed, possibly after supernova explosions, forming a bound system of two compact objects (neutron stars or black holes) that inspiral on each other as a result of the emission of GWs. The number of massive systems that remain bounded after two supernova explosions (or prompt core-collapse) is uncertain, as is the time to coalescence (the delay), or intrinsic parameters such as masses, spins, and eccentricity, which depend on a complicated evolution scenario involving a common-envelope and mass-transfer \citep{2002ApJ...572..407B,2008ApJS..174..223B}.
The distance probed with actual interferometers is about 30-40 Mpc for NSNSs \citep{2010arXiv1003.2481T,2012arXiv1203.2674T}, but the next generation of detectors such as Ad~LIGO/Virgo \footnote{http://www.ligo.caltech.edu/docs/M/M060056-10.pdf, https://wwwcascina.virgo.infn.it/advirgo/docs.html} should be taking data with a sensitivity approximately ten times greater, pushing the horizon up to about 450 Mpc \citep{2010CQGra..27q3001A}. With the third-generation interferometer Einstein Telescope (ET) \citep{2010CQGra..27s4002P}, the horizon of compact binaries is expected to reach cosmological distances of $z>2-3$, where it may become possible to study the evolution of the sources over redshift~\footnote{http://www.et-gw.eu/etdsdocument}. 
By horizon we mean the distance of an NSNS binary whose GW signal would exceed the detector threshold, assuming an optimal orientation and localization of the source.
In addition to the emission produced by the coalescence of the individually resolved binary systems, the superposition of a large number of unresolved sources at high redshifts will produce a background of gravitational waves (see \citet{2011RAA....11..369R,2011ApJ...739...86Z,2011PhRvD..84l4037M, 2011PhRvD..84h4004R,2012PhRvD..85j4024W,2012PhRvD..86l2001R,2012arXiv1209.0595Z}, for the most recent studies and also \citet{2001MNRAS.324..797S,2003MNRAS.346.1197F,2009PhRvD..79f2002R}) that may dominate the cosmological background in the range $10-1000$ Hz where terrestrial detectors are the most sensitive. In the Laser Interferometer Space Antenna (LISA) band between $10^{-4}-0.1$ Hz \citep{1998AAS...193.4803B}, it is expected that background from extragalactic sources will be detectable after substraction of the galactic foreground from white dwarf binaries.

In this paper we focus on sources that are not members of any dense environment such as globular clusters. Additional formation channels of BHBH in globular clusters have been investigated by \citet{2007PhRvD..76f1504O}, \citet{2008ApJ...676.1162S}, and their dependence on metallicity has been explored by \citet{2014MNRAS.441.3703Z}.

Recent studies have shown that the properties of compact object binaries 
strongly depend on the metallicity of the stellar population. The formation rate of 
binaries containing black holes sharply increases with decreasing metallicity, as indicated both 
by observations of binares containg a massive BH accreting from a Wolf-Rayet star \citep{2010MmSAI..81..302B,2011ApJ...730..140B},
and by binary population synthesis \citep{2010ApJ...714.1217B}. Moreover, it has been shown that at low metallicity the
typical mass of a black hole increases \citep{2010ApJ...715L.138B}. While for the 
solar metallicity the highest mass of a black hole is 
about $10-15\,M_\odot$, for a metallicity of about 10\% solar value
it rises to above $30\, M_\odot$. 
The  high-redshift population of mergers of compact objects originates in populations
of stars with various metallicities, including very low ones. Previous studies of \citet{2004ApJ...615L..65S,2008ApJ...677..813M, 2013MNRAS.433.1094S}, and many others
have shown that in the very early Universe, where the environment was nearly metal free, binaries might have been created.
A investigation of the gravitational wave background must therefore take the effects of metallicity into account.

We studied the effect of the metallicity on the GW background from compact binaries (NSNS, BHNS or BHBH). In Sect. 2 we describe how we use the simulation code {\tt StarTrack}, in Sect. 3 we discuss  the calculation of the coalescence rate as a function of redshift, in Sect. 4 we derive the GW background and describe the numerical simulations, in Sect. 5 we present our results, and finally in Sect. 6 we summarize our main conclusions.

\section{Simulating a population of compact binaries with {\tt StarTrack}}
The parameters that characterize the simulated compact binary systems were generated using the binary evolution code {\tt StarTrack}. They include the
masses of the two components $M_1$ and $M_2$, the time delay, $t_d$, and the eccentricity of the orbit $e$.
The masses of the compact objects are determined by the masses of the two progenitors and by mass loss and mass exchange during the common-envelope phase (CE), when the most massive star, already a neutron star or a black hole, accretes mass from its companion, a red giant.
The delay is the sum of the evolution time between the birth of the two massive progenitors and the formation of the compact binary ($t_{evol}$), and the merging time, or the time it takes for the two stars to coalesce through the emission of GWs ($t_{mr}$) \citep{1963PhRv..131..435P},
\begin{equation}
t_d=t_{evol}+t_{mr}.
\end{equation}

To model the population of compact object binaries,
we used the {\tt StarTrack} population synthesis code \citep{2002ApJ...572..407B} to perform a suite
of Monte Carlo simulations of the stellar evolution of stars in environments
of two typical metallicities: $Z=Z_{\odot}=0.02$ and $Z=10\% \, Z_{\odot}=0.002$ 
\citep{2010ApJ...715L.138B}.
In these calculations we employed the recent estimates of mass-loss rates 
\citep{2010ApJ...714.1217B}. They include revised mass-loss for 
O/B stars following \citet{2001A&A...369..574V} and Wolf-Rayet stars investigated by \citet{1998A&A...335.1003H} and \citet{2005A&A...442..587V}. Additionally, the luminous
blue variable mass-loss rates were calibrated so as to account for the formation of BH with $15M_{\odot}$ in a solar metallicity environment
\citep{2011ApJ...742...84O}. This also results in the formation of BH with $30M_\odot$ in subsolar metallicities like the BH in IC10 X-1 (Z$=0.3Z_\odot$)
\citep{2007ApJ...669L..21P}, \citep{2008ApJ...678L..17S}.
We calculated a population of two million massive binary stars, 
tracking the ensuing formation of relativistic
binary compact objects: double neutron stars (NSNS), double black hole binaries 
(BHBH), and mixed systems (BHNS). We used updated stellar and 
binary physics, including results from supernova simulations
and compact object formation \citep{2012ApJ...749...91F}, incorporating elaborate 
mechanisms for treating stellar interactions such as mass-transfer episodes \citep{2008ApJS..174..223B} 
or tidal synchronization and circularization \citep{1981A&A....99..126H}. We focused on the common-envelope evolution phase \citep{1984ApJ...277..355W}, which is
crucial for close double compact object formation because the associadted mass-transfer 
allows for an efficient hardening of the binary \citep{2012ApJ...759...52D}. 

An important parameter describing the common-envelope is the $\lambda$ coefficient.
It is a measure of how strongly the donor envelope is bound to the core.
We used the realistic calculations of the common-envelope coefficient $\lambda$ 
performed by \citet{2010ApJ...716..114X}, which now depends on the evolutionary stage of the donor, 
its mass at the zero age main sequence (ZAMS), the mass of its envelope, and its radius.
The orbital contraction occurring during the common-envelope can be
sufficiently efficient to cause the individual stars in the binary to coalesce
and form a single highly rotating object, thereby aborting any further binary 
evolution and preventing the formation of a double compact object. Because of 
significant radial expansion, stars crossing the Hertzsprung gap (HG) very 
frequently initiate a common-envelope phase. 
HG stars do not have a clear 
entropy jump at the core-envelope transition \citep{2004ApJ...601.1058I}; if such a 
star overflows its Roche lobe and initiates a common-envelope phase, the 
inspiralling is expected to lead to a coalescence \citep{2000ARA&A..38..113T}. In 
particular, it has been estimated that for a solar metallicity environment (e.g., 
our Galaxy), properly accounting for the HG gap may lead to a reduction in the 
merger rates of BHBH binaries by $\sim 2-3$ orders of magnitude \citep{2007ApJ...662..504B}.
In contrast, in a low-metallicity environment this suppression is much less
severe ($\sim 1$ order of magnitude; \citet{2010ApJ...715L.138B}).
A growing presence of metals increases the opacity of the medium they occupy. The increased radiation 
pressure of trapped photons then causes the medium to expand. Therefore, stars with a low amount of metals do not grow 
during their evolution as much as high-metallicity stars. This makes them less likely to overfill their Roche lobes and initiate a CE on the HG.
The details of the common-envelope phase are not yet fully understood, 
and thus in what follows we consider two set of models, one that does not take into 
account the suppression (optimistic models: marked with A), and a set that assumes 
the highest possible suppression (pessimistic models: marked with B).
For NSs we adopt natal kick distributions from observations of single 
Galactic pulsars \citep{2005MNRAS.360..974H} with $\sigma =265$~km/s. 
However, for BHs we draw kicks from the same distribution but with a lower
magnitude: inverse proportional to the amount of fall-back expected at the BH 
formation \citep{2001ApJ...554..548F}. In particular, for most massive BHs
that form with the full fall-back (the direct BH formation), the magnitude of the
natal kick is zero (more details on modeling can be found in \citet{2012ApJ...759...52D}).

The detailed list of models considered in this paper is presented in Table~\ref{tab:Models}.
To study how metallicity will affect the gravitational-wave background, we considered three
different cases. Two of them are of single metallicity, where all stars were born in the same environment 
(solar metallicity marked as Z or $10\%$ of solar metallicity denoted by z). In the third case we assume that half of all stars
have solar metallicity, the other half have $10\%$ of solar metallicity (denoted by $50/50$ in Table \ref{tab:Models}.)
This choice is meant to provide a first-order
assessment of the dependence on metallicity of the GW background (see Sect. \ref{sec:results} for a more detailed disscution).

Detailed properties such as average total mass or average "chirp mass" are
listed in Table \ref{tab:Properties}.

\begin{table}[ht]
\caption{List of stellar evolution models. HG specifies whether we have an optimistic (+) or pessimistic (-) model. For more details see text.}
\label{tab:Models}
\begin{minipage}{\linewidth}\centering
\begin{tabular}{c c c }
\hline\hline
Model & Metallicity & HG \\
\hline
AZ & $Z_{\odot}$ & \multirow{3}{*}{+} \\
Az & $10\%$ $Z_{\odot}$ & \\
A & $50/50$ ($Z_{\odot}$ and $10\%$ $Z_{\odot}$) &  \\
\hline
BZ & $Z_{\odot}$ & \multirow{3}{*}{-} \\
Bz & $10\%$ $Z_{\odot}$ &  \\
B & $50/50$ ($Z_{\odot}$ and $10\%$ $Z_{\odot}$) & \\
\hline
\end{tabular}
\end{minipage}
\end{table}

\begin{table}
\caption{Statistical properties of compact binaries used in the simulations of single metallicity populations. For each model
we list the average total mass of a binary, the average "chirp mass", and the average frequency at the last stable orbit.}
\label{tab:Properties}
\begin{minipage}{\linewidth}\centering
\begin{tabular}{c c c c}
\hline\hline
Model & $<\textrm{M}_{tot}>$ [M$_\odot$] & $<M_{chirp}>$ [M$_\odot$] & $<f_{lso}>$ [Hz]\\
\hline\hline
\multicolumn{4}{c}{NSNS}\\
\hline
AZ & 2.43 & 1.05 & 1809.71 \\
Az & 2.51 & 1.09 & 1756.74 \\
A & 2.45 & 1.06 & 1794.74\\
BZ & 2.43 & 1.05 & 1811.26 \\
Bz & 2.49 & 1.08 & 1768.23 \\
B & 2.44 & 1.06 & 1802.96\\
\hline
\multicolumn{4}{c}{BHNS}\\
\hline
AZ & 9.91 & 3.17 & 444.89 \\
Az & 11.66 & 3.17 & 398.82 \\
A & 11.17 & 3.18 & 412.00\\
BZ & 9.85 & 3.13 & 448.39 \\
Bz & 12.45 & 3.21 & 371.55 \\
B & 12.21 & 3.20 & 378.43\\
\hline
\multicolumn{4}{c}{BHBH}\\
\hline
AZ & 15.56 & 6.74 & 283.66 \\
Az & 30.31 & 13.08 & 188.09 \\
A & 28.85 & 12.45 & 197.79\\
BZ & 15.60 & 6.76 & 282.59 \\
Bz & 22.41 & 9.54 & 215.14 \\
B & 21.78 & 9.28 & 221.31\\
\hline
\end{tabular}
\end{minipage}
\end{table}

\section{Cosmic coalescence rate}
We assumed that for each category of binaries ($j$=NSNS, BHNS, BHBH), the coalescence rate tracks the star formation rate (SFR), albeit with some delay $t_d$ from formating the massive binary to the final merger: 
\begin{equation}
\dot{\rho}_c^j(z) = A^j \int \frac{\dot{\rho}_*(z_f)}{1+z_f}P(t_d)\,dt_d,
\label{eq:rateV}
\end{equation}
where $\dot{\rho}_*$ is the star formation rate in  M$_\odot$Mpc$^{-3}$yr$^{-1}$ , $A^j$ is the mass fraction of the progenitors in M$_{\odot}^{-1}$ 
(see appendix for more details), $P^j(t_d)$ is the probability distribution of the delay $t_d$, $z_f$ is the redshift at which the progenitor binary forms, and $z$ is the redshift 
at which the two compact objects coalesce. The factor $(1+z_f)^{-1}$ converts the rate in the source 
frame into a rate in the observer frame. 
The redshifts $z_f$ and $z$ are related by the delay time $t_d$,
which is the
difference in lookback times between $z_f$ and $z$:
\begin{equation}
t_d = \frac{1}{H_0}
\int_z^{z_f} \frac{dz'}{(1 + z')E(\Omega, z')}\,.
\end{equation}
The coalescence rate per redshift bin is then given by
\begin{equation}
\frac{dR^j}{dz}(z)= \dot{\rho}_c(z) \frac{dV}{dz}(z)\,.
\label{eq:rate}
\end{equation}
The co-moving volume element is
\begin{equation}
\frac{dV}{dz}(z)=4 \pi \frac{c}{H_0} \frac{r(z)^2}{E(\Omega,z)}\,,
\end{equation}
where
\begin{equation}
r(z)= \frac{c}{H_0}\int_0^z \frac{dz'}{E(\Omega, z')}\,,
\label{eq:distance}
\end{equation}
and 
\begin{equation}
E(\Omega,z)=\sqrt{\Omega_{\Lambda}+\Omega_{m}(1+z)^3}\,.
\end{equation}

\begin{figure}
\centering
\includegraphics[angle=000,width=\columnwidth]{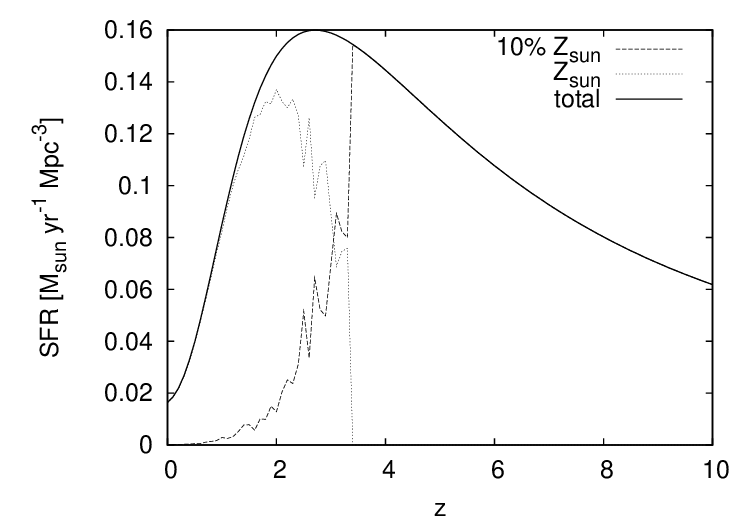}
\caption
{Star formation rate as a function of redshift. The dashed line corresponds to the low-metallicity environment, the dotted line represents solar metallicity, and
the solid line shows total SFR.}
\label{fig:sfr}
\end{figure}

In these simulations, we used the cosmological model derived from seven years of WMAP observations \citep{2011ApJS..192...18K} and assumed a flat Universe with $\Omega_m=0.27$,
$\Omega_{\Lambda}=0.73$ and Hubble parameter $H_0=70.3$\,km\, s$^{-1}$\,Mpc$^{-1}$.

We used the SFR of \citet{2004ApJ...613..200S}, which was adopted by \citet{2013ApJ...779...72D} to include influence of metallicity.
They have considered two different scenarios for the metallicity evolution denoted as {\it slow}, where coefficients of parametric equation for SFR are provided by \citet{1999ApJ...522..604P}, and
{\it fast}, with coefficients taken from \citet{2007ApJ...670..584Y}. For each galaxy the authors combined the {\it slow} and {\it fast} scenario by taking an average of two.
The metallicity evolution obtained in this way does not agree with observations of the local Universe. Therefore, the authors used two normalization to fit their model to the existing observations.
In the first case, labeled {\it high-end}, the metallicity is divided by a factor of $1.7$, which corresponds to the upper limit of the metallicity at redshift $z \sim 0$, while in the second case, labeled
{\it low-end}, the metallicity is divided by $3$ to match the SDSS observations. For more details see Sect. 2 of a paper by \citet{2013ApJ...779...72D}. In this paper, we considered the high-end scenario.

The adopted SFR is shown in Fig.~\ref{fig:sfr}.
The dashed line shows the star formation rate for a low-metallicity environment (that includes groups III and IV shown in Fig. 2 of \citet{2013ApJ...779...72D}). From that distribution all stars with metallicity $Z=10\% Z_\odot$ were chosen.
The dotted line represents the solar metallicity (groups I and II shown in Fig. 2 of \citet{2013ApJ...779...72D}), which we adopted for the higher metallicity case.
In the third case, where half of the stars have lower metallicity and other half solar metallicity, we used both these distributions.

Account for uncertainties in the SFR, we also considered the model presented by \citet{2003MNRAS.339..312S}.
However, as was already noticed by \citet{2011RAA....11..369R}, because of the uncertainties on the star formation history essentially affect redshift $z>2$ where sources do not contribute significantly to the integrated signal, our final results are not expected to vary by more than a factor 2, depending on the choice of the SFR. Therefore, we focused on one particular SFR choice, which is shown in Fig.~\ref{fig:sfr}.

The final rate is the sum of the rates from the different categories:
\begin{equation}
\frac{dR}{dz}(z)=\frac{dR^{\mathrm{NSNS}}}{dz}(z)+\frac{dR^{\mathrm{BHNS}}}{dz}(z)+\frac{dR^{\mathrm{BHBH}}}{dz}(z).
\end{equation}

By integrating this equation over redshift, we calculated the expected number of events per year for all considered models. A detailed list of all values is shown in Table \ref{tab:Rates}.
In addition, it shows the contributions of different types of binaries (NSNS, BHNS and BHBH) to the total coalescence rate.
As expected, models labeled A predict significantly more sources because they were initially more numerous. The percentage of NSNS decreases with the decrease in metallicity.
Models with solar metalicity are dominated by NSNS, while models with lower metallicity are dominated by BHBH.

\begin{table}[ht]
\caption{Total coalescence rate of all compact binaries per year (Col. 2).
In the last three columns we show the contribution of each type of compact binaries to the total coalescence rate.}
\label{tab:Rates}
\begin{minipage}{\linewidth}\centering
\begin{tabular}{c c c c c}
\hline\hline
Model & rate [$yr^{-1}$] & rate$_{\mathrm{NSNS}}$ [$\%$] & rate$_{\mathrm{BHNS}}$ [$\%$] & rate$_{\mathrm{BHBH}}$ [$\%$] \\
\hline
AZ & 622 572 & 71.62 & 3.70 & 24.69 \\
Az & 1 606 240 & 10.62 & 3.56 & 85.82 \\
A & 1 264 605 & 27.55 & 3.54 & 68.91 \\
BZ & 154 929 & 84.78 & 2.09 & 13.13 \\
Bz & 319 304 & 10.70 & 11.50 & 77.80 \\
B & 267 677 & 34.52 & 8.31 & 57.16 \\
\hline
\end{tabular}
\end{minipage}
\end{table}

\section{GW background}
The superposition of the GW signal from sources at all redshifts creates a background whose 
spectrum is usually
characterized by the dimensionless energy density parameter \citep{1999PhRvD..59j2001A}:
\begin{equation}
\Omega_{gw}(f)=\frac{1}{\rho_c}\frac{d\rho_{gw}}{d\ln f},
\end{equation}
where $\rho_{\rm gw}$ is the gravitational energy density and $\rho_c=\frac{3c^2H_0^2}{8 \pi G}$
is the critical energy density needed to make the Universe flat today.
The GW spectrum from the population of extragalactic binaries is given by the expression 
\begin{equation}
\Omega_{gw}(f)=\frac{1}{\rho_c c} f F(f),
\label{eq:omega_flux}
\end{equation}
where $F(f)$ is the total flux and $f$ is the observed frequency.
We estimated the total flux (in erg Hz$^{-1}$) as the sum of the individual contributions
whose redshift and masses are given by the {\tt StarTrack} simulations:  
\begin{equation}
F(f)= T^{-1} \sum_{k=1}^{N} \frac{1}{4 \pi d_L^2(z^k)} \frac{dE_{gw}}{df}(f,M^k,\mathcal{M}_c^k,z^k),
\label{eq:flux}
\end{equation}
where $\frac{dE_{gw}}{df} (f,M^k,\mathcal{M}_c^k,z^k)$ is the spectral energy density (in erg Hz$^{-1}$s$^{-1}$) averaged over the orientation for a binary at redshift $z^k$, with total mass $M^k$ and "chirp mass" $\mathcal{M}_c^k$. $N$ is the number of sources produced by {\tt StarTrack} and corresponds to the number of coalescences in $T=1$ yr.
That number is derived from merger rate estimates. The normalization factor $T^{-1}$ ensures that the flux has the correct dimension.
The evolution of coalescing binaries is comprised of three phases, the inspiral, the merger, and the ring-down of the final black hole. The early inspiral phase and the ring-down can be modeled accurately by post-Newtonian expansion and perturbation theory, while our knowledge of the late inspiral and the merger rely on numerical relativity. 

We followed \citet{2011ApJ...739...86Z,2012arXiv1209.0595Z} and used:
1) Taylor T4 templates with 3.0 post-Newtonian corrections for the population of NSNS. The merger and ring-down carry much uncertainty
due to the lack of precise knowledge of the amount of energy emitted in these 
phases. In a recent paper, \citet{2012arXiv1209.0595Z} have calculated the contribution of an additional Gaussian post-merger emission centered on 2 kHz and concluded that it only affected the high-frequency part of the background, whose contribution to the signal-to-noise ratio (S/N) is negligible.
2) We also used the phenomenological templates of \citet{2008PhRvD..77j4017A}, including inspiral, merger, and ring-down, constructed by combining analytical and numerical relativity results for a nonspinning BHBH \footnote{Templates exist also for nonprecessing spinning BHs \citep{2011PhRvL.106x1101A}. Using these templates would have very little effect on the results \citet{2011ApJ...739...86Z} and would not change our conclusions.}. \citet{2009PhRvD..79d4030S} have shown that BHNS signals with a mass ratio higher than 6 look very similar to BHBH waveforms because no tidal disruption occurs outside the last stable orbit. 

In the early inspiral phase, the spectral energy density can be described by the quadrupole approximation  (assuming circular orbit) \footnote{We used the formula of \citet{1963PhRv..131..435P} for nonzero eccentricity (the initial eccentricity is provided by {\tt StarTrack} ), but it has no effect on the frequency band of terrestrial detectors where most of the orbits have been circularized.}:
\begin{eqnarray}
\frac{dE_{gw}}{df} (f,f_{lso},\mathcal{M}_c,z)= \ \nonumber
\frac{(G \pi)^{2/3} (\mathcal{M}_c(1+z))^{5/3}}{3} f^{-1/3},\\ \nonumber
\ \mathrm{for}\,\ f<f_{lso}/(1+z),
\end{eqnarray}
where $f_{lso}=c^3/(6\sqrt{6}\pi GM)$ is the GW frequency at the last stable orbit (LSO).

For each category of binary, we can use the following approximate analytical expression, which involves the rates derived in the previous section and average quantities :
\begin{equation}
F^j(f) \simeq \int_0^{z_{max}} \frac{1}{4 \pi d_L^2(z)} \frac{dE_{gw}}{df}(f,\bar{f_{lso}}^j,\bar{\mathcal{M}_c}^j,z) \frac{dR^j}{dz}(z)dz,
\label{eq:flux_anal}
\end{equation} 
where $j=$ NSNS, BHNS or BHBH, $\bar{f_{lso}}^j$ is the average frequency at the LSO, and $\bar{\mathcal{M}_c}^j$ is the average "chirp mass" of the population.
Inserting Eq.~\ref{eq:flux_anal} into Eq.~\ref{eq:omega_flux}, we obtain

\begin{eqnarray}
\Omega^j_{gw}(f) \simeq & \frac{\displaystyle 2 \pi^{2/3}G^{5/3}} {\displaystyle 9c^3 H_0^2 } (\bar{\mathcal{M}}_c^j)^{5/3}  f^{2/3}\\ \nonumber
&\displaystyle \int_0^{z_{sup}^j(f)}  \frac{dR^j}{\displaystyle dz}(z) \frac{dz}{\displaystyle r(z)(1+z)^{1/3}} dz,
\label{eq:omega}
\end{eqnarray}
with
\begin{equation}
z_{sup}^j (f)=
\left\lbrace
\begin{array}{l}
z_{max} \,\  \mathrm{ if} f < \bar{f_{lso}}^j/(1+z_{max})\\
(\bar{f_{lso}}^j / f)-1 \,\   \mathrm{otherwise}.\\
\end{array}
\right.
\label{eq-zsup}
\end{equation}
This expression provides a very good estimate of the background in the low-frequency region (before the LSO) and when the "chirp mass" and the delay distributions show little correlation.

\section{Results and discussion}
\label{sec:results}
\begin{figure}
\centering
\includegraphics[angle=000,width=\columnwidth]{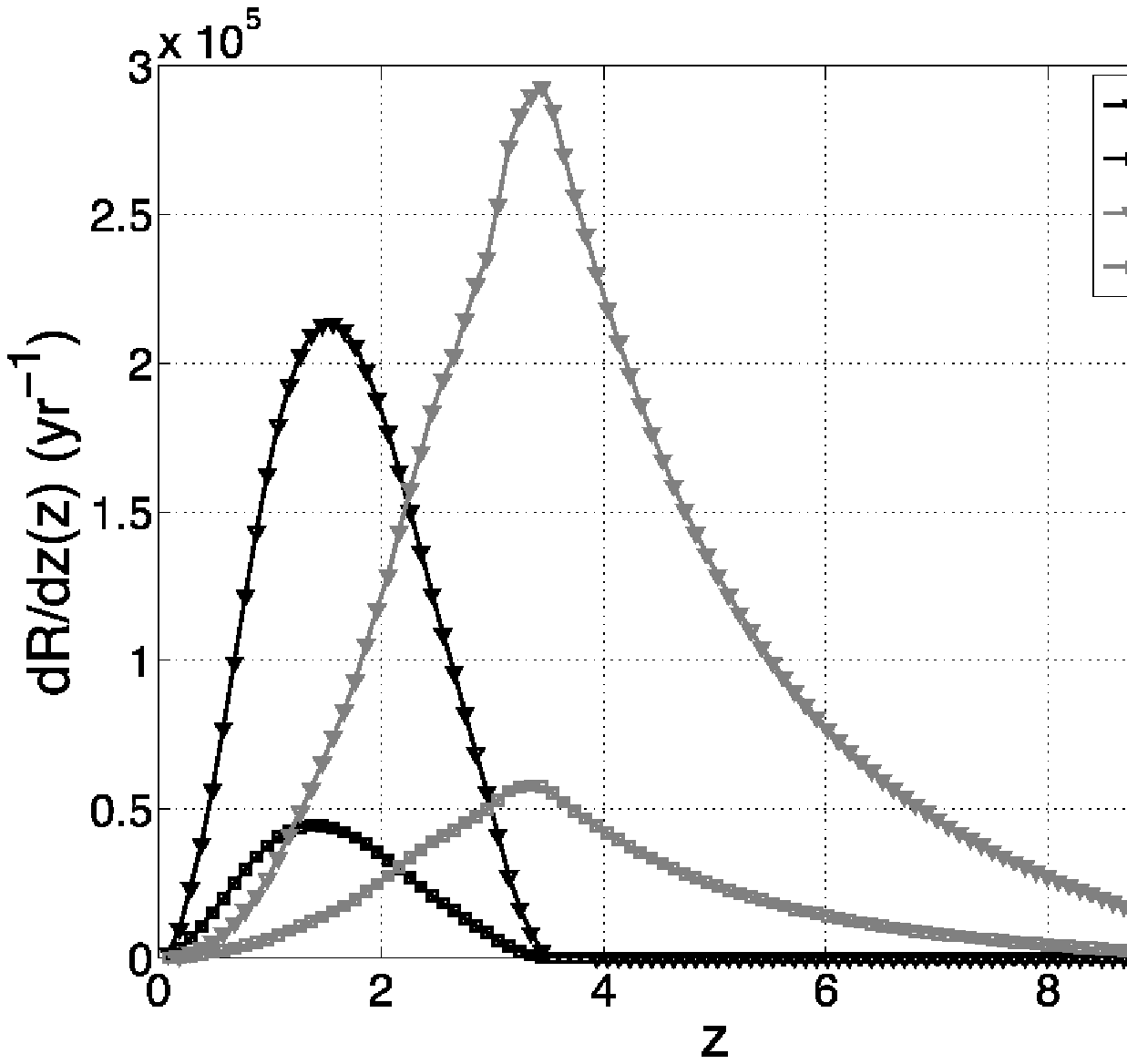}
\caption{
Rate of compact binary coalescences as a function of the redshift for the different metallicity values and for the different models Table~\ref{tab:Models}:
AZ (black continuous line with triangle symbol), BZ (black continuous line with square symbol), Az (gray continuous line with triangle symbol), Bz (gray continuous line with square symbol).
Black lines correspond to high (solar) metallicity and gray lines to low metallicity, triangles represent optimistic models (A), squares pessimistic ones (B).
\label{fig-dRdz}}
\end{figure}

We present the dependence of compact object binary
coalescence rate on the redshift in Fig.~\ref{fig-dRdz}.
The shape of the distributions is very similar for all the models within a single metallicity, as the distribution of the delay does not show noticeable differences from one model to the other, expect for differences in the lowest  delay value, which is much lower than the Hubble time and does not significantly affect the final distribution. The coalescence rate peaks at around $z \sim 1.5$ for solar metallicity and around $z \sim 3.2$ for lower metallicity. It is shifted toward lower redshift than the star formation rate, where the peak for solar metallicity is around $z \sim 2$ and is around $z \sim 3.6$ for lower metallicity.
For comaparison see Fig.~\ref{fig:sfr}.
The normalization, on the other hand, changes because it is linked to the compact object 
formation rate specific to each model. Typically, 
models A (optimistic) correspond to higher rates than models 
B (pessimistic), and models with low metallicity 
lead to higher overall rates than the models with 
high metallicity.
We must note that the models we present were calculated for a
few particular population synthesis models. Thus the absolute normalization 
of the results is uncertain, similar to any rate calculation 
in the literature. 
It is also clearly visible that the obtained normalization is more affected by details of the HG treatment (distinction between A and B models) than by any metallicity effect.
Therefore, improving stellar evolution models will be crucial in determining the true value of merger rates.
A detailed discussion of the expected rates based on the {\tt StarTrack} population synthesis models has been published separately by \citet{2012ApJ...759...52D}.

\begin{figure*}
\centering
\includegraphics[angle=270,width=0.65\textwidth]{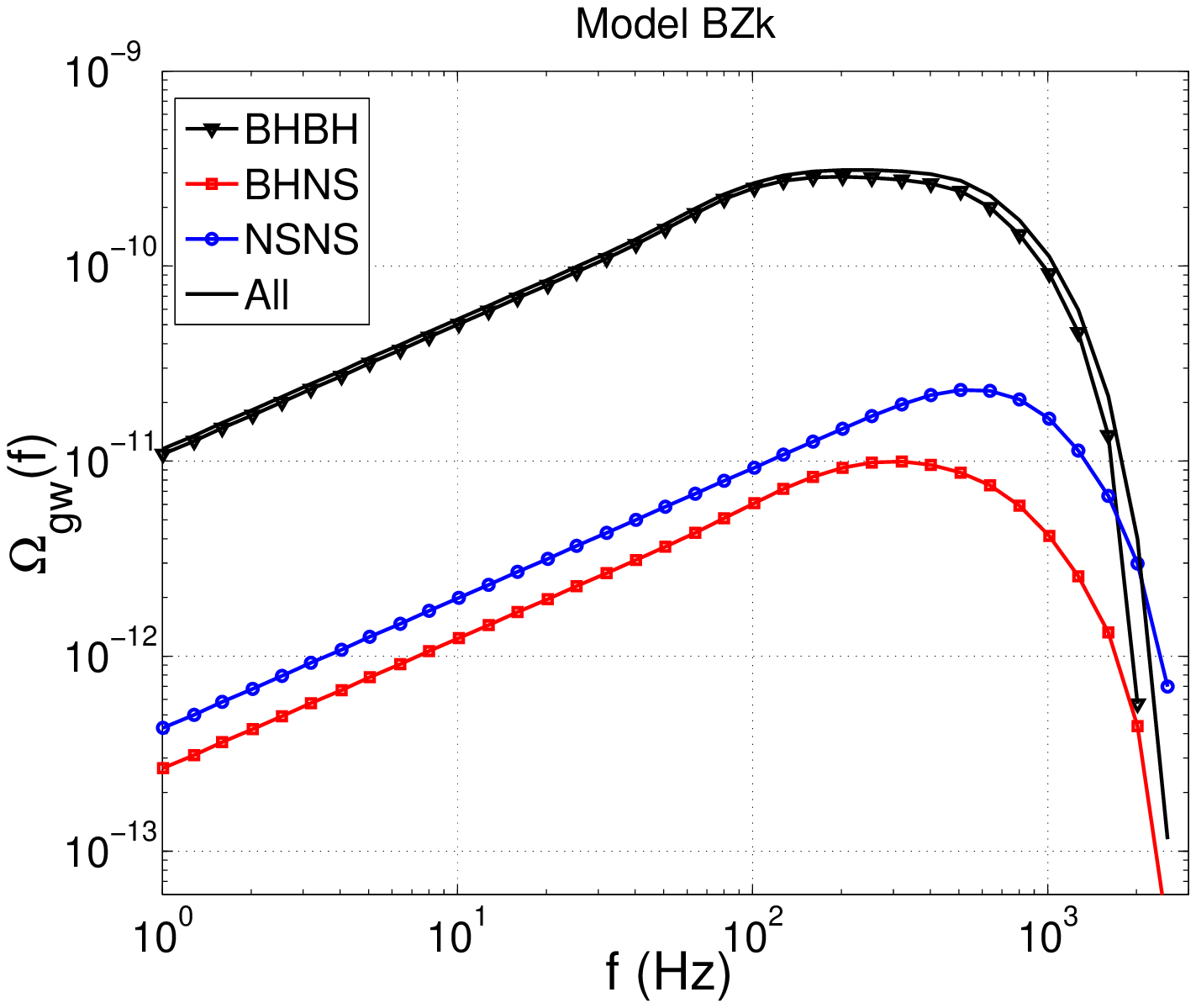}
\caption{
Energy density parameter of the background produced by the coalescence
of the different type of binaries: two neutron stars (blue continuous line with circle symbol), a neutron star and a black hole (red continuous line with square symbol) two black holes (black continuous line with triangle symbol), for model B with high metallicity $Z=Z_{\odot}$. The black line with no symbol corresponds to the sum of all the contributions.}
\label{fig-omega_B}
\end{figure*}

The resulting gravitational-wave background spectrum for the model BZ is presented in Fig.~\ref{fig-omega_B} with the individual contributions of 
each type of binaries shown separately. 

For BHBH and BHNS,  $\Omega_{gw}$ increases as $f^{2/3}$ (inspiral phase), then as $f^{5/3}$ (merger phase), before it reaches a maximum and decreases dramatically. The slope change and the peak occur at frequencies corresponding roughly to the LSO and the end of the ring-down phase at $z \sim 1.5$ where the coalescence rate is highest.
For NSNS, the slope drops from $2/3$ to $1/3$ at around $\sim 100$ Hz, as predicted by the TaylorT4 formula. Adding a Gaussian post-merger emission would enhance the NSNS contribution to between $\sim 0.5-2.5$ kHz (see \citet{2012arXiv1209.0595Z}), but would have no effect on the detectability of the background, as we show in the next section.  
The amplitude of the background scales up with the rate and the average "chirp mass" (see Eq.~\ref{eq:omega}) and is larger for the BHBH background than for the BHNS and NSNS contributions by $1-3$ orders of magnitude. Consequently, at low frequencies, the background is dominated by the signal from BHBH. After the BHBH peak, the spectrum decreases dramatically, but the signal from the NSNS is buried even for high frequencies. The population of BHNS contributes very little to the total background since it is masked by the BHBH signal at all frequencies.

\begin{figure*}
\centering
\includegraphics[angle=000,width=1\textwidth]{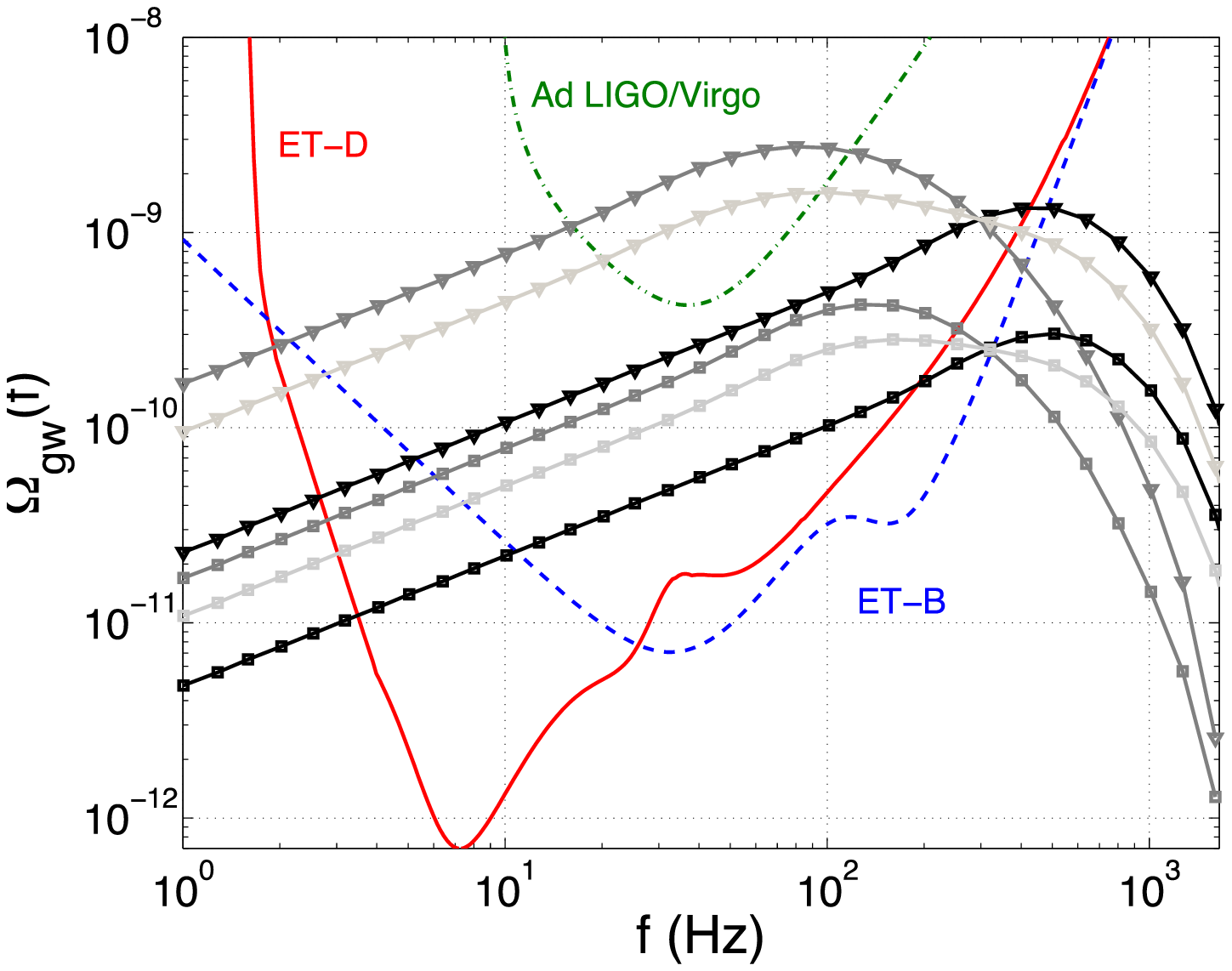}
\caption{
Energy density parameter of the background produced by the coalescence 
of compact binaries for the different values of the metallicity and for the different models considered in this paper:
AZ (black continuous line with triangle symbol), BZ (black continuous line with square symbol), Az (gray continuous line with triangle symbol), Bz (gray continuous line with square symbol), A (light-gray continuous line with triangle symbol), and B (light-gray continuous line with square symbol).
Black lines correspond to high (solar) metallicity, gray lines to low metallicity, and light-gray to populations with mixed metallicity.
The sensitivity of ET-B, ET-D and advanced LIGO/Virgo are also shown for comparison purposes.}
\label{fig-omega}
\end{figure*}

The spectra for the different models of Table~\ref{tab:Models} are presented together in Fig.~\ref{fig-omega}. 
The effect  of the choice of
model A or B (with a different treatment of the CE with HG donor)
mainly affects the normalization of the spectrum.
The three upper lines with triangle symbols correspond to the optimistic model A, while the three lines below them with square symbols correspond to the pessimistic model B. Within each group we show
the metallicity.
The change of the metallicity of the underlying population 
changes both the amplitude and shape of the spectrum.
The lower metallicity population contains more
high-mass BHs, and therefore the low-frequency peak shifts
to the lower frequencies. 
The population with a single low metallicity produces the highest level of background, the population with single high metallicity is the one with the lowest background, while for mixed metallicites the population background
lies in the middle. The peaks of each model are connected to the corresponding redshift dependence of the merger rates. Models with low metallicity (gray lines) have peaks around $f \sim 100$ Hz, models with high metallicity (black lines) have their maximum around
$f \sim 400$ Hz, while mixed population models (light gray lines) smoothly connect both peaks. In this last case we observe a break in the spectrum around $f \sim 100$ Hz followed by a gentle plateau. 
The amplitude of the spectrum shifts up with decreasing metallicity. This is due to an increase in the 
formation rate of compact object binaries with decreasing metallicity.

The gravitational-wave background as presented in Fig.~\ref{fig-omega} is highest in the frequency band of terrestrial interferometers ($1$\,Hz-$10$\,kHz) and might be accessible by the next generations of detectors.
The standard detection method of cross-correlation between two detectors gives an S/N
\begin{equation}
S/N =\frac{3 H_0^2}{4 \pi^2} \sqrt{2T} \left[ \int_0^\infty
df \frac{\gamma^2(f)\Omega_{\rm gw}^2(f)}{f^6 P_1(f)P_2(f)} \right]^{1/2},
\label{eq:snrCC}
\end{equation}
where $\gamma$ is the overlap reduction function characterizing the loss in sensitivity due to the separation and the relative orientation of the detectors, normalized so that it is one for L-shaped co-aligned and co-located detectors, $P_1$ and $P_2$ 
are the strain noise power spectral densities of the two detectors. 
Notice that this method is optimal for a Gaussian stochastic background, which is not the case of the backgrounds calculated in this paper (the sources do not overlap in the frequency range of terrestrial detectors). However, the ET Mock Data Challenges \citep{2012PhRvD..86l2001R} has proven that such non-Gaussian backgrounds can still be detected with no loss of sensitivity by the cross-correlation statistic.

We considered a pair of co-aligned and co-located advanced (second-generation detectors such as the LIGO Hanford pair in the US.
A similar sensitivity can be obtained by combining the future network of detector pairs worldwide (LIGO Hanford and LIGO Livingston in the US, Virgo in Italy, GEO in Germany, IndIGO in India, KAGRA in Japan), the number of pairs balancing the loss in sensitivity caused by their separation and relative orientation (\citet{2011ApJ...739...86Z}, Mandic private communication).
We also considered a pair of V-shaped detectors (an angle of $\pi/3$ between the two arms) separated by an angle of $2\pi/3$, with the sensitivity of the third-generation triangle detector ET. We used two models for the noise, ET-B \citep{2008arXiv0810.0604H}, which is about ten times better than the sensitivity of advanced LIGO/Virgo with a low-frequency cutoff at $1-3$ Hz, and ET-D \citep{2011CQGra..28i4013H} with a significant improvement at frequencies below 30 Hz. 

\begin{figure}
\centering
\includegraphics[angle=270,width=0.5\textwidth]{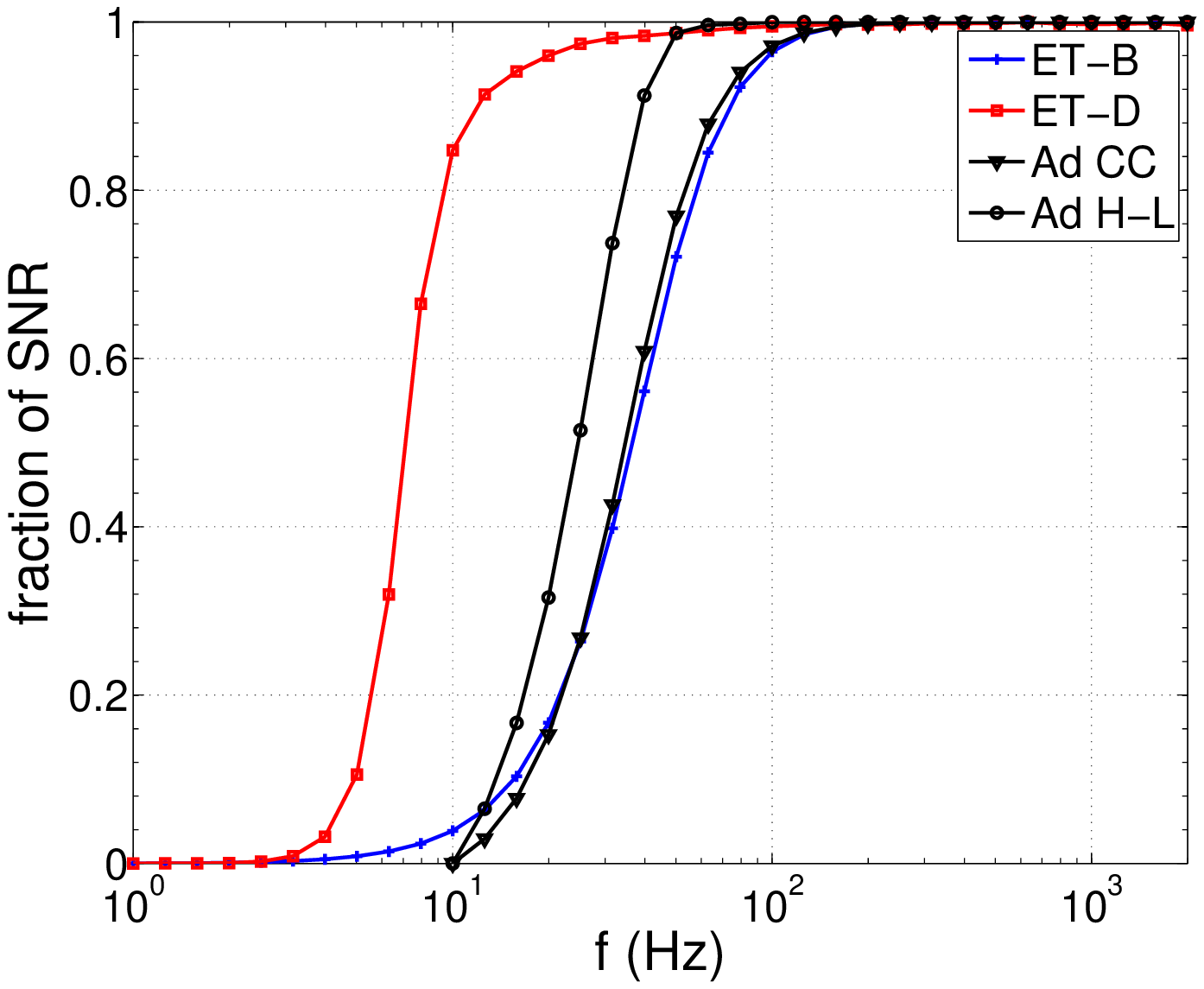}
\caption{Contribution to the S/N of frequencies $<f$ for co-aligned and co-located advanced detectors, separated advanced detectors (the LIGO Hanford/Livingston pair), and the third-generation Einstein Telescope (ET-B and ET-D).}
\label{fig-snr_fraction}
\end{figure}

In Fig.~\ref{fig-snr_fraction}, we show the cumulative S/N as a function of the frequency for advanced and ET detectors for the model BZ (considering other models or  a simple power law $\Omega_{gw} \sim f^{2/3}$ does not change the results significantly). Clearly, most of the S/N comes from a relatively small frequency
interval, which depends on the configuration pair, but extends from a few Hz to $<$ 100 Hz. At these frequencies, BHBH contribution dominates the background signal.

The S/N is reported in Table ~\ref{tab:SNR} for all the models, assuming an integration time of 1 yr. 
Models with low metallicities should be detected by advanced detectors with $\textrm{S/N} \sim 0.7-7$, so only an optimistic model will exceed the detection threshold of $\textrm{S/N}_{thr}=2.5$ (for a false-alarm rate and
false-dismissal rate of 10\% \citep{1999PhRvD..59j2001A}). Models with high metallicity are beyond reach by the factor of a few. For populations with mixed metallicites only an optimistic model (A) is accessible, while a pessimistic model (B) is below the detection treshold.
With the Einstein Telescope, all the models should be detectable with a very high signal-to-noise ratio of between $\textrm{S/N} \sim 13-471$ for ET-B and $24-855$ for ET-D.

The detection threshold was calculated by using the following equation \citep{1999PhRvD..59j2001A}:
\begin{equation}
\textrm{S/N}_{thr} = \sqrt{2} (erfc^{-1}(2 \alpha)-erfc^{-1}(2 \beta)),
\end{equation}
where $erfc^{-1}$ is the inverse complementary error function, $\alpha$ the false-alarm rate, and $\gamma$ the detection rate (i.e, 1- false-dismissal rate).

\begin{table}[ht]
\caption{Signal-to-noise ratio obtained by cross-correlating two co-aligned and co-located Advanced detectors or a pair of two ET detectors with the sensitivity ET-B or ET-D for the different models presented in Table~\ref{tab:Models}, }
\label{tab:SNR}
\begin{minipage}{\linewidth}\centering
\begin{tabular}{c c c c}
\hline\hline
Model & Adv CC & ET-B & ET-D \\
\hline
AZ & 0.925 & 61.782 & 116.683\\
Az & 7.138 & 471.502 & 854.678\\
A & 4.003 & 264.726 & 482.754\\
BZ & 0.192 & 12.811 & 24.202\\
Bz & 0.710 & 47.780 & 86.195\\
B & 0.444 & 29.868 & 54.487\\
\hline
\end{tabular}
\end{minipage}
\end{table}

Another promising detection method that has been tested in the context of the ET Mock Data Challenge \citep{2012PhRvD..86l2001R} for the population of NSNS is to examine the residual of the null stream, which contains no GW signal and arises from the sum of the three ET detectors.
However, \citet{2014PhRvD..89h4063M} has shown that non-Gaussian backgrounds can be detected with no loss of sensitivity by the standard cross-correlation
statistic used for Gaussian backgrounds as long as there is a strong enough signal present in the observation time, even if they do not overlap. 
This results was verified by the standard cross-correlation used in the Einstein Telescope Mock Data Challenges \citep{2012PhRvD..86l2001R, 2014PhRvD..89h4046R}.

\section{Conclusions}
We investigated the effect of the metallicity on the background from extragalactic compact binary coalescences. We used the {\tt StarTrack}  population synthesis code to perform a suite
of Monte Carlo simulations of the stellar evolution of binary systems in environments
of two typical metallicities: $Z=Z_{\odot}=0.02$ and $Z=10\% \, Z_{\odot}=0.002$, and a population containing stars with both metallicites.
By using these three simple models, we calculated the effect of metallicity on our analysis. We chose two extreme cases in the local Universe (all stars have solar metallicity or all stars have $10 \%$ of solar metallicity) and one case that lies between these extreme values (a population where half of the stars have solar metallicity an the other half have $10 \%$ of solar metallicity).
The real metallicity distribution is far more complicated and depends on redshift. It was also shown that the observed metallicity distribution is correlated with the stellar mass and with
SFR in a very complicated way \citep{2010MNRAS.408.2115M}. Therefore, our results should be treated as a first step toward more complex models that will better describe realistic stellar populations.

We found that for all models, the background is dominated by the binary black-hole contribution in the frequency range where terrestrial detectors are the most sensitive, with a common structure: an evolution as $\Omega_{gw} \sim f^{2/3}$ corresponding to the inspiral phase, then as $\Omega_{gw} \sim f^{5/3} $ corresponding to the merger phase, a maximum at $0.6-1$ kHz, and finally at high frequencies, the final stages of the binary neutron star contribution in the post-Newtonian regime with $\Omega_{gw} \sim f^{1/3}.$
A high metallicity $Z=Z_{\odot}=0.02$ produces less massive system, thus the amplitude of the spectrum is lower and the peak is located in a high-frequence region. In fact, when we conisidered populations consisting solely of solar metallicity, we will not be able to observe that peak, because it will be outside the sensitivity window of terrestial interferometers. By adding more stars with lower metallicity to the population, we observed that the maximum of the spectrum shifted toward lower frequencies (compare black, light gray, and gray lines in Fig.~\ref{fig-omega}). In the frequency range between $100$ Hz and $500$ Hz the model that contains both metallicites has form of a power law.
The slope of that part of the spectrum strongly depends on the metallicity distribution. Therefore, by observing where gravitational-wave background will break
and the slope of received spectrum, we are able to deduce metallicity distribution of the considered populations. By taking into account a realistic metallicity distribution model one should obtain
a gravitational-wave background that lies somewhere between two extreme cases, as shown in Fig.~\ref{fig-omega}.

All our models are detectable by the Einstein Telescope after a year of observations with a high signal-to-noise ratio (between $13-855$). Optimistic models with a low metallicity are also detectable by advanced LIGO/Virgo, while models with solar metallicity are below the detection threshold, which is consistent with the estimate of detectability by \citet{2012PhRvD..85j4024W}, \citet{2011ApJ...739...86Z}, \citet{2011PhRvD..84l4037M}, and \citet{2012arXiv1209.0595Z}. 
Since the background from compact binary coalescences has a good chance of being detected in the near future with the second-generation of detectors, realistic estimates of its shape and amplitude are crucial to optimize the search and later interpret the results. The LIGO/Virgo stochastic group and the ET Science Team have started Mock Data Challenges based on the models presented in this paper, whose results are presented by \citet{2014PhRvD..89h4046R}.

\section*{Acknowledgments}
{
This work was supported by the Polish grants DEC-2011/01/V/ST9/03171, DEC-2011/01/V/ST9/4071, and DEC-2013/01/ASPERA/ST9/00001
of the Associated European Laboratory ``Astrophysics Poland-France''.
We are grateful to Xingjiang Zhu for kindly providing a matlab code to calculate the TaylorT4 waveforms for the NSNS population.
}

\bibliography{ecc}

\appendix
\section[]{Normalization of the cosmic coalescence rate}
To normalize the coalescence rate, we need to take into account various factors. First of all, our simulations cannot be treated as a complete sample
of realistic stars. We do not cover the whole parameter space, in particular, the mass range is narrower than in reality. Moreover, not all stars that are born in binaries
will end their lives as a compact binaries. Most of them will not survive the common-envelope stage.

The coalescence rate for one particular binary $i$ is given by the following formula:
\begin{equation}
\label{rate}
\dot{\rho}_{c,i}(z)=\frac{1}{N_{sim}}n_{bin}f_s,
\end{equation}

where $N_{sim}$ is the total number of binaries that started their evolution on the ZAMS in the {\tt StarTrack} code. In our case that number is $2\times 10^{6}$,
$n_{bin}$ is the number density of stars in binaries, $f_s$ is the fraction of simulated binaries.

Not only binares are formed in the Universe. The total star formation rate ($\dot{\rho_*}$) is the sum
of the contribution from binaries and single stars,

\begin{equation}
\label{sfr}
\dot{\rho_*}=\dot{\rho_*}_{bin}+\dot{\rho_*}_{sin} \, [M_\odot Mpc^{-3} yr^{-1}].
\end{equation}

The number density is defined as

\begin{equation}
\label{n}
n_k=\frac{\dot{\rho_*}_k}{<M_k>} \, [Mpc^{-3} yr^{-1}],
\end{equation}
where k="bin" or "sin" and $<M_k>$ is the average mass of a star in a binary, if $k=bin$, or a single star, if $k=sin$.

Another important parameter is binary fraction, which is defined as a fraction between the number of binary systems and the total number
of all systems (binary and single),

\begin{equation}
\label{fb}
f_b=\frac{N_{bin}}{N_{bin}+N_{sin}}=\frac{0.5n_{bin}}{0.5n_{bin}+n_{sin}},
\end{equation} 

where $N_{bin}$ is the number of binaries, not stars, so we need a factor of 0.5 in the second par of the equation.
We set the binary fraction to be $f_b=0.5$, as for a current stellar populations.

By transforming Eq. \ref{fb}, we obtain number density of a single stars

\begin{equation}
n_{sin}=\frac{1}{2}n_{bin}\frac{1-f_b}{f_b},
\end{equation}

then, combing Eqs. \ref{sfr} and \ref{n}, we have the number density of stars in binaries,

\begin{eqnarray}
\label{nbin}
n_{bin}=\dot{\rho_*} \frac{2}{\frac{1-f_b}{f_b}<M_{sin}>+2<M_{bin}>},\\
n'_{bin}=\frac{2}{\frac{1-f_b}{f_b}<M_{sin}>+2<M_{bin}>}.
\end{eqnarray}

As a next step we need to compute the average masses of a single star and of those that are components of a binary.
We used initial mass function (IMF) proposed by \citet{1955ApJ...121..161S} and a flat mass-ratio distribution,

\begin{eqnarray}
\Psi(M)&=&KM^{-2.7},\\
\Phi(q)&=&2q,
\end{eqnarray}

where $K^{-1}=\int_{M_{min}}^{M_{max}} M^{-2.7} dM$, $M_{min}=0.08M_\odot$, $M_{max}=150M_\odot$.
%$K=0.0232 M_\odot^1.7$.

With the IMF we can compute the  average masses of single stars and of the components of binaries,
\begin{equation}
<M_{sin}>=\int_{M_{min}}^{M_{max}} M\Psi(M) dM, % = 0.1933 M_\odot
\end{equation}

\begin{equation}
<M_{bin}>=\frac{1}{2}\int_{M_{min}}^{M_{max}} M\Psi(M) dM \int_0^1 dq (1+q) \Phi(q). %=0.1611M_\odot
\end{equation}

The form of $f_s$ is a consequence of the specific way of drawing masses of the components in {\tt StarTrack}.
First, the mass of the primary is drawn from the Salpeter distribution with $M_{min,sim}=5M_\odot$ and $M_{max,sim}=M_{max}$. 
Then, the mass of the secondary is determined by the mass ratio, which is taken from a flat distribution.
The mass of the second component has to be greater than $3M_\odot$. The fraction of simulated stars can be expressed by

\begin{equation}
f_s=\frac{\int_{M_{min,sim}}^{M_{max}} dM \Psi(M)\int_{q_{min}(M)}^1 dq \Phi(q)}{\int_{M_{min}}^{M_{max}} dM \Psi(M)\int_0^1 dq \Phi(q)},
\end{equation}

where $q_{min}(M)=\frac{3}{M}$.

The normalization $A^k$ (k=NSNS, BHNS or BHBH) can be defined as

\begin{equation}
A^k=\frac{N^k}{N_{sim}}n'_{bin} f_s \, [M_\odot^{-1}].
\end{equation}

$N^k$ is the number of k-type binaries in our simulations. 

\end{document}